\documentclass[notitlepage,amsmath,amssymb,aps]{revtex4-1}

\usepackage{graphicx}
\usepackage{dcolumn}
\usepackage{bm}

\begin{document}
\title{Time and Thermodynamics\\Extended Discussion on ``Time \& clocks: A thermodynamic approach"}
\author{Atanu Chatterjee}
\email{achatterjee3@wpi.edu} 
\author{Germano Iannacchione}
\email{gsiannac@wpi.edu}
\affiliation{Department of Physics\\
Worcester Polytechnic Institute\\
100 Institute Road, Worcester, MA 01605, USA}

\begin{abstract}
In the paper, ``Time \& clocks: A thermodynamic approach" Lucia and Grisolia describe the connections between the physical nature of time and macroscopic irreversibility in thermodynamics. They also discuss the possibility of constructing a thermodynamic clock that links the two, through an approach based on the behaviour of a black body. Their work primarily focuses on the macroscopic irreversibilty, and their attempt to define `thermodynamic time' is grounded on the concepts from non-equilibrium thermodynamics such as, entropy generation and time-dependent thermodynamic fluxes - in and out of the system. In this letter, we present a first principles approach based on the Maupertuis principle to describe the connection between time and thermodynamics for equilibrium phenomena. Our novel interpretation of the temperature as a functional allows us to extend our formalism to irreversible processes when the local equilibrium hypothesis is satisfied. Through our framework we also establish a functional relationship between equilibrium and non-equilibrium time-scales while distinguishing between the two. 
\end{abstract}

\maketitle

It is certainly true that the notion of time has a direction associated with it. The natural ordering of events in our physical reality allows us to distinguish the past from the present, and we perceive this directionality in the flow of time as an irreversible phenomenon. Therefore, the concept of time in a process that is reversible is not only irrelevant, but may seem physically absurd. In their work, Lucia and Grisolia address the connection between time and thermodynamic irreversibility from a first principles approach~\cite{lucia2020time,lucia2008probability,lucia2012maximum}. The authors define a unit of `thermodynamic time' as the ratio of the local entropy to the rate of local entropy production. Since, this definition of time involves entropy production, it assumes a meaningful interpretation only during irreversible processes. This is physically relevant in the context of the second law of thermodynamics. Therefore, the authors claim, that it is not possible to realise reversible clocks in nature is logically sound. We, however take a different approach in this letter. Since, the basis of thermodynamics lies on equilibrium phenomena, we feel that an interpretation of time in equilibrium thermodynamics should act as the stepping stone for the physically relevant far-from-equilibrium scenario~\cite{lucia2008probability,chatterjee2016thermodynamics,georgiev2016road}. 

A classical physical system can be described by a set of canonical coordinates, $(p, q)$ where the coordinates can be indexed to represent degrees of freedom and $\dot{q}=p/m$. These generalized system of coordinates denote the position $(q)$ and momentum $(p)$ of a particle (or a collection of particles) in the phase-space. The Hamiltonian of such a system corresponds to its total energy and is given by, $H(p,q) = p^2/2m + V(q)$ (or the sum of the kinetic energy and potential energy). Whereas, the Lagrangian is written as the difference between the kinetic and potential energies, $L(q,\dot{q}) = p^2/2m - V(q)$. One can define two types of action integrals in classical mechanics: the Maupertuis' action or abbreviated action, $\mathcal{S}_0$ and the Eulerian action, $\mathcal{S}$ given by, 
\begin{equation}
    \mathcal{S}_0 = \int p\mathrm{d}q\quad\text{and}\quad\mathcal{S} = \int L\mathrm{d}t
    \label{eqn1}
\end{equation}
Further, the Hamiltonian and the Lagrangian can be related to each other by the following Legendre transformation, $p\dot{q} - H(p,q) = L(q,\dot{q})$.
One can rewrite the Legendre form above by substituting the abbreviated action from Equation~\ref{eqn1} as,
\begin{equation}
    \int p\mathrm{d}q = \int L\mathrm{d}t + \int H\mathrm{d}t\quad\text{and}\quad\int_{q_1}^{q_2}\mathrm{d}\mathcal{S}_0 = \int_{t_1}^{t_2}\mathrm{d}t(L + H)\equiv\mathcal{S}_0 = \mathcal{S} + \tau H
    \label{eqn2}
\end{equation}
For a conservative system driven between two states $q(t_1) = q_1$ and $q(t_2) = q_2$, the Hamiltonian $H(p,q)$ is conserved in time. Moreover, Equation~\ref{eqn2} denotes the kinetic energy, and the two action functionals differ by a constant, $\tau H$ where $\tau = t_2 - t_1$, or the time spent by the system along a reversible path connecting two equilibrium states. Since, both Maupertuis and Eulerian action describe the same physical process, they achieve a stationary value simultaneously. Conversely, Equation~\ref{eqn2} in differential forms yields, $\delta\mathcal{S}_0 = \delta\mathcal{S} + H\delta\tau + \tau\delta H\equiv\delta\mathcal{S}_0 = \delta\mathcal{S}$, since $\tau$ and $H$ are constants (therefore, their variation is zero). The Hamilton-Jacobi equation relating Eulerian action with the Hamiltonian can be written as,
\begin{equation}
    \frac{\partial\mathcal{S}}{\partial t} + H = 0
    \label{eqn3}
\end{equation}
We had discussed before the total variation in the Eulerian action along a constant energy path connecting two equilibrium states with a characteristic time, `$\tau$' as $\delta\mathcal{S} = \delta\mathcal{S}_0 - H\delta\tau - \tau\delta H$; and also from Equation~\ref{eqn3} we have $\delta\mathcal{S} + H\delta t = 0$. Therefore, we can conclude
\begin{equation}
    \frac{\partial\mathcal{S}_0}{\partial H} = \tau    
    \label{eqn4}
\end{equation}
On integrating Equation~\ref{eqn4} we get, $\mathcal{S}_0 - \tau H = 0$. The above set of Equations~\ref{eqn1}-\ref{eqn4} presents the Maupertuis' derivation of the equations of motion for a classical (conservative) system and they hold true iff, the canonical momentum can be expressed in terms of the Lagrangian as follows, $p_i = \partial L(q,\dot q)/\partial\dot q$~\cite{landau2013course}. For a system driven along a reversible cycle, the first law of thermodynamics reduces to $\delta Q - \delta W = 0$, where $Q$ denotes the heat given to the system and $W$ denotes the work done by the system. Since, the system is driven along a reversible cycle, $\oint \mathrm{d}U = 0$, and $\tau$ denotes the characteristic time of the cycle. As the system is conservative, the total energy of the system is equal to the heat given to the system therefore, $\mathcal{S}_0 - \tau H \Rightarrow \mathcal{S}_0 - \tau Q = 0$. From the equipartition theorem we know, $Q = Nfk_BT$ where $N$ is the number of particles, $f$ the degree of freedom, $k_B$ the Boltzmann's constant and $T$ the canonical temperature. Therefore, we can rewrite the equipartition theorem in terms of the abbreviated action, the canonical temperature, and cycle time as follows,
\begin{equation}
    \mathcal{S}_0 - \tau Q = \mathcal{S}_0 - \tau (Nfk_BT)\quad\text{or}\quad T = \frac{\omega\mathcal{S}_0}{Nfk_B}
    \label{eqn5}
\end{equation}
In the above equation, $\omega = 1/\tau$ denotes the cycle frequency. The above approach which has been previously discussed by Garc{\'i}a-Morales et. al., proposes temperature as a functional on the energy landscape as compared to the conventional state-point interpretation~\cite{garcia2008thermodynamics}. When the action is minimized, i.e. $\delta\mathcal{S}_0 = 0\Rightarrow\delta T = 0$, implies that isotherms in the phase-space are unique paths along which temperature remains constant. Moreover, for a system quasi-statically driven between a pair of phase-space coordinate, we can calculate the spatially averaged momenta of the micro-states as follows,
\begin{equation}
    \langle p\rangle_q = \frac{\int_{q_1}^{q_2}p\mathrm{d}q}{\int_{q_1}^{q_2}\mathrm{d}q} = \lambda\mathcal{S}_0
    \label{eqn6}
\end{equation}
The integral of the position element between $q_2$ and $q_1$ is $\Delta q = q_2 - q_1 \equiv \ell$ and $\lambda = 1/\ell$ (where $\ell$ is the characteristic length). Therefore, Equation~\ref{eqn5} can be also written as,
\begin{equation}
    T = \frac{\omega\mathcal{S}_0}{Nfk_B} = \frac{\langle p\rangle_q}{Nfk_B}\frac{\ell}{\tau} = \frac{\langle p\rangle_q\cdot\langle\overrightarrow{u}\rangle}{Nfk_B} = \frac{2\langle K\rangle}{Nfk_B}
    \label{eqn7}
\end{equation}
Where $\langle K\rangle$ is the mean equilibrium kinetic energy, and the ratio $\ell/\tau$ is the mean characteristic velocity, $\langle\overrightarrow{u}\rangle$ that describes the rate at which the quasi-static process takes place. The above result is in agreement with the kinetic theory of gases. The Boltzmann's factor, $\beta = 1/k_BT$ can be written as, $Nf/(\langle p\rangle_q\cdot\langle\overrightarrow{u}\rangle)$, which for an ideal gas becomes, $3N/m\langle u\rangle^2$ a well-known result, see for instance~\cite{pauli2000thermodynamics}.

Let us consider a simple example where a piston with mass $(m)$ compresses a gas confined inside a cylinder (with cross-sectional area, $A$) quasi-statically and then allows it to expand. The process is cyclic as $q(t_1) = q(t_2)$, and the micro-states ($e_i$) distributed inside the piston-cylinder system is given by the Boltzmann's distribution, $\rho(e_i) = \exp(-\beta e_i)/Z$ ($Z$ is the partition function). The work done by the piston-cylinder during compression is given by,
\begin{equation}
    W = - P\Delta V = - \int_1^2 P\mathrm{d}V = -\frac{1}{A}\int_1^2 \dot{p}(A\mathrm{d}q) = - \langle\overrightarrow{u}\rangle\int_1^2\mathrm{d}p
    \label{eqn8}
\end{equation}
In the above equation, we use the Newton's law of motion and rewrite pressure as $P = F/A = \dot p/A$. From the ideal gas equation we have, $P\Delta V = Nfk_B\Delta T$. Therefore, Equation~\ref{eqn8} can be written as, $Nfk_B\Delta T = \langle\overrightarrow{u}\rangle\Delta p$ which further reduces for to $\langle T\rangle = (\langle\overrightarrow{u}\rangle\cdot\langle p\rangle_q)/Nfk_B$ where $\langle T\rangle$ is the canonical temperature of the piston-cylinder system which is in agreement with the kinetic theory. In the above description we assume reversible expansion (compression) of an ideal gas. For an irreversible process (neglecting gravity), the total force exerted by the pressurized gas can be modeled as a damped oscillator, such that $\dot p \rightarrow \dot p + \gamma (p/m)^n$ where $\gamma (p/m)^n$ is the viscous drag as the piston oscillates. For $n=1$, the work done during irreversible compression becomes $- \langle\overrightarrow{u}\rangle\int_1^2\mathrm{d}p - (\gamma/m)\int_1^2p\mathrm{d}q = - \langle\overrightarrow{u}\rangle\cdot\langle\overrightarrow{p}\rangle_q - (\gamma/mA)\langle\overrightarrow{p}\rangle_q\Delta V_{12}$. Due to the presence of this additional drag component, the irreversible time-scale will be much larger than the reversible/equilibrium time-scale.

We can also calculate the statistical entropy, $S$ in terms of the abbreviated action. From the Maxwell's relations we have, $1/T = (\partial S/\partial H)_{N,V}$ and from Equation~\ref{eqn4}, $(\partial\mathcal{S}_0/\partial H) = \tau$. From Equations~\ref{eqn4} and~\ref{eqn5} we have,
\begin{equation}
    Nfk_B\frac{\partial\ln\mathcal{S}_0}{\partial H} = \frac{Nfk_B}{\mathcal{S}_0}\frac{\partial\mathcal{S}_0}{\partial H} = Nfk_B\frac{\tau}{\mathcal{S}_0} = \frac{1}{T} = \frac{\partial S}{\partial H}
    \label{eqn9}
\end{equation}
Thus, we can conclude that $S = Nfk_B\ln\mathcal{S}_0 = k\ln\mathcal{S}_0$ where the constant $Nfk_B$ is denoted by $k$. For the case of thermodynamic equilibrium, we obtain the second law of thermodynamics as a consequence of the principle of stationary action, or $\delta\mathcal{S}_0 = 0\equiv\delta S = 0$. Let us now consider the non-equilibrium scenario and calculate the rate of entropy production.
\begin{equation}
    \dot S\equiv\frac{\mathrm{d}S}{\mathrm{d}t} =  \frac{k}{\mathcal{S}_0}\frac{\partial\mathcal{S}_0}{\partial t} = \frac{\omega}{T}\frac{\partial\mathcal{S}_0}{\partial t}
    \label{eqn10}
\end{equation}
\begin{figure*}
    \centering
    \includegraphics{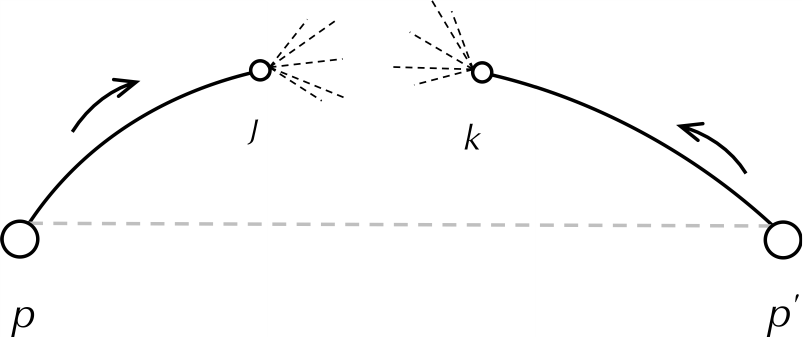}
    \caption{Figure presents a scenario where two stationary states, $p$ and $p^\prime$ are connected by a reversible path (light grey, dotted) and an irreversible path (black, solid). Along the irreversible path, a segment, $jk$ is identified which can be traversed by numerous irreversible paths and a single reversible path (straight line connecting $j$ to $k$).}
    \label{fig}
\end{figure*}
The authors present a measure of thermodynamic time, $\tau_{neq} = S/\dot S$, which in our formulation takes the following form,
\begin{equation}
    \tau_{neq} = \frac{S}{\dot S} = \frac{TS}{\omega}\cdot\Big(\frac{\partial\mathcal{S}_0}{\partial t}\Big)^{-1} = \tau\Big(\frac{TS}{W}\Big)_{neq} = \frac{\tau}{\eta_{neq}}
    \label{eqn11}
\end{equation}
In the above equation, $TS$ denotes the non-equilibrium heat dissipation, $(\partial\mathcal{S}_0/\partial t)$ (from Hamilton-Jacobi equation) represents non-equilibrium work (lost), and $\tau$ represents the time spent along a reversible path. The ratio of the non-equilibrium work to the non-equilibrium heat dissipation is the non-equilibrium efficiency, $\eta_{neq}$. The non-equilibrium efficiency, $\eta_{neq}<1$, therefore $\tau_{neq}>\tau$. Since, the reversible path between end states follow the shortest trajectory $(\tau)$, any other trajectory will therefore be longer and hence would take longer time $(\tau_{neq})$, as seen earlier during irreversible compression/expansion in a piston-cylinder system. We consider the case when $\dot S = 0$, i.e. no local entropy generation and the system is at equilibrium. In this case, both $\tau_{neq}$ and $\tau$ become undefined because time loses its physical significance and therefore temporal symmetry is preserved, which is one way of stating that the system is conservative and the process is reversible. When the system is out-of-equilibrium, i.e. dissipative or $\dot S>0$ the thermodynamic variables become undefined. Therefore, in the absence of a definition for entropy in such a system it is difficult to interpret the relationship between clock-time and thermodynamic time. We, however have a special case: the non-equilibrium steady-state where the average of the local entropy fluctuations vanishes. In this state, the ratio of the probabilities of the forward to the backward trajectories diverge exponentially as a function of $\dot S$~\cite{taniguchi2007onsager}. A special scenario for a system at non-equilibrium steady-state is to satisfy the local equilibrium hypothesis, for instance in a non-turbulent Rayleigh-B{\'e}nard convection~\cite{chatterjee2019coexisting,yadati2019spatio}. The nature of the time-scale for this scenario is discussed in detail below. 

Let an irreversible path connect two stationary states $(p,p^\prime)$ in the phase-space, a scenario described in Figure~\ref{fig}. We consider a segment with intermediate states, $j$ and $k$ along this path. The rate at which entropy is generated while moving from $j$ to $k$ is denoted by,
\begin{equation}
    \frac{\mathrm{d}S}{\mathrm{d}t} = \frac{(S_j-S_p) - (S_k-S_{p^\prime})}{t_j - t_k} = \frac{S_j-S_k}{t_j-t_k} + \frac{S_{p^\prime}-S_p}{t_j - t_k} = \frac{k\ln\mathcal{S}_{0_{jk}} - k\ln\mathcal{S}_{0_{kj}}}{\tau_{jk}} + \frac{C}{\tau_{jk}}
    \label{eqn12}
\end{equation}
The terms, $S_j$ and $S_k$ denote the entropy at the states $j$ (in relative to $p$) and $k$ (in relative to $p^\prime$); the terms $\mathcal{S}_{0_{jk}}=\int_j^kp\mathrm{d}q$ and $\mathcal{S}_{0_{kj}}=\int_k^jp\mathrm{d}q$ denote the magnitude of action along the segment $j$ to $k$ and $k$ to $j$ respectively. Finally, $\tau_{jk}$ denotes the time interval between points $j$ and $k$ along the shortest path. The difference between the entropies of the two stationary end states, $S_{p^\prime} - S_p$ is a constant. Rearranging the constant term, $C/\tau_{jk}$ by rewriting it as $k\ln\alpha/\tau_{jk}$, $(C=k\ln\alpha)$ and substituting it back in Equation~\ref{eqn12} gives us,
\begin{equation}
    \dot S\equiv\frac{\mathrm{d}S}{\mathrm{d}t} = \frac{k\ln\mathcal{S}_{0_{jk}} - k\ln\mathcal{S}_{0_{kj}}}{\tau_{jk}} + \frac{k\ln\alpha}{\tau_{jk}}\Rightarrow \frac{\mathcal{S}_{0_{jk}}}{\mathcal{S}_{0_{kj}}} = \frac{1}{\alpha}\exp(\tau_{jk}\dot S/k)
    \label{eqn13}
\end{equation}
When $\dot S > 0$, the magnitude of the action $\mathcal{S}_{0_{kj}}=(\mathcal{S}_{0_{jk}}/\alpha)\exp(-\tau_{jk}\dot S/k)$ decays exponentially as a function of $\dot S$. Thus, the more probable path is from $k$ to $j$ as it minimizes the action through dissipation (since $\dot S>0$). Equation~\ref{eqn13} can be generalized for a scenario where between a pair of stationary states, $(p,p^\prime)$ there is a single reversible path with (minimal) action $\mathcal{S}_0$ and a collection of irreversible paths each of which generates entropy, $\dot{S}_i$~\cite{taniguchi2007onsager,el2015least}. 
\begin{equation}
    \mathcal{S}_0 = \frac{\mathcal{S}_{0_i}}{\alpha}\exp(-t\dot S_i/k)
    \label{eqn14}
\end{equation}
The above ratio, $\mathcal{S}_0/\mathcal{S}_{0_i}\approx\tau/\tau_i$ dictates the likelihood of an irreversible path connecting two stationary states. Looking at the relationship, one can conclude that the time taken to realize an irreversible process connecting one equilibrium state to another grows exponentially when compared to a reversible path connecting the two. Therefore in many physical systems that are driven out-of-equilibrium, stationary states are absent as the system is always at a state of dynamic equilibrium. This results in the spontaneous creation of metastable states. While these states are not strictly stationary (due to entropy generation and presence of fluxes), they do however show characteristics similar to an equilibrium stationary state~\cite{vilar2001thermodynamics,srinivasarao2019biologically}. The out-of-equilibrium system thus locks itself in these local equilibrium states while dissipating heat into the surrounding and giving rise to emergent structural complexities, for example, the Rayeligh-B{\'e}nard convection. Under such circumstances the spatial variation of the entropy and temperature are crucial which we have discussed in our previous works~\cite{chatterjee2019many,chatterjee2019coexisting,yadati2019spatio,chatterjee2019overview}. 

Finally, the authors discuss the consequence of relativistic effect on temperature. For $v<<c$ where $c$ is the velocity of light, length and time transform as, $x^\prime = \gamma x$ and $t^\prime = \gamma t$ where $(x^\prime, t^\prime)$ are the coordinates in $K^\prime$ frame, and $(x,t)$ in the $K$ frame~\cite{landau2013course}. From our definition of temperature in Equations~\ref{eqn5} and~\ref{eqn7}, we can find that temperature is a function of action, frequency, and energy. Since, action is an invariant under Lorentz transformation, therefore entropy is a Lorentz invariant. Further, temperature should satisfy the Lorentz transformation as, $T^\prime = T/\gamma$, which is in agreement with the authors proposal~\cite{lucia2020time,farias2017temperature,marevs2010relativistic}. 
%


\begin{thebibliography}{18}%
\makeatletter
\providecommand \@ifxundefined [1]{%
 \@ifx{#1\undefined}
}%
\providecommand \@ifnum [1]{%
 \ifnum #1\expandafter \@firstoftwo
 \else \expandafter \@secondoftwo
 \fi
}%
\providecommand \@ifx [1]{%
 \ifx #1\expandafter \@firstoftwo
 \else \expandafter \@secondoftwo
 \fi
}%
\providecommand \natexlab [1]{#1}%
\providecommand \enquote  [1]{``#1''}%
\providecommand \bibnamefont  [1]{#1}%
\providecommand \bibfnamefont [1]{#1}%
\providecommand \citenamefont [1]{#1}%
\providecommand \href@noop [0]{\@secondoftwo}%
\providecommand \href [0]{\begingroup \@sanitize@url \@href}%
\providecommand \@href[1]{\@@startlink{#1}\@@href}%
\providecommand \@@href[1]{\endgroup#1\@@endlink}%
\providecommand \@sanitize@url [0]{\catcode `\\12\catcode `\$12\catcode
  `\&12\catcode `\#12\catcode `\^12\catcode `\_12\catcode `\%12\relax}%
\providecommand \@@startlink[1]{}%
\providecommand \@@endlink[0]{}%
\providecommand \url  [0]{\begingroup\@sanitize@url \@url }%
\providecommand \@url [1]{\endgroup\@href {#1}{\urlprefix }}%
\providecommand \urlprefix  [0]{URL }%
\providecommand \Eprint [0]{\href }%
\providecommand \doibase [0]{https://doi.org/}%
\providecommand \selectlanguage [0]{\@gobble}%
\providecommand \bibinfo  [0]{\@secondoftwo}%
\providecommand \bibfield  [0]{\@secondoftwo}%
\providecommand \translation [1]{[#1]}%
\providecommand \BibitemOpen [0]{}%
\providecommand \bibitemStop [0]{}%
\providecommand \bibitemNoStop [0]{.\EOS\space}%
\providecommand \EOS [0]{\spacefactor3000\relax}%
\providecommand \BibitemShut  [1]{\csname bibitem#1\endcsname}%
\let\auto@bib@innerbib\@empty
\bibitem [{\citenamefont {Lucia}\ and\ \citenamefont
  {Grisolia}(2020)}]{lucia2020time}%
  \BibitemOpen
  \bibfield  {author} {\bibinfo {author} {\bibfnamefont {U.}~\bibnamefont
  {Lucia}}\ and\ \bibinfo {author} {\bibfnamefont {G.}~\bibnamefont
  {Grisolia}},\ }\bibfield  {title} {\bibinfo {title} {Time \& clocks: A
  thermodynamic approach},\ }\href@noop {} {\bibfield  {journal} {\bibinfo
  {journal} {Results in Physics}\ }\textbf {\bibinfo {volume} {16}},\ \bibinfo
  {pages} {102977} (\bibinfo {year} {2020})}\BibitemShut {NoStop}%
\bibitem [{\citenamefont {Lucia}(2008)}]{lucia2008probability}%
  \BibitemOpen
  \bibfield  {author} {\bibinfo {author} {\bibfnamefont {U.}~\bibnamefont
  {Lucia}},\ }\bibfield  {title} {\bibinfo {title} {Probability, ergodicity,
  irreversibility and dynamical systems},\ }\bibfield  {booktitle} {\emph
  {\bibinfo {booktitle} {Proceedings of the Royal Society of London A:
  Mathematical, Physical and Engineering Sciences}},\ }\href@noop {} {\ \textbf
  {\bibinfo {volume} {464}},\ \bibinfo {pages} {1089} (\bibinfo {year}
  {2008})}\BibitemShut {NoStop}%
\bibitem [{\citenamefont {Lucia}(2012)}]{lucia2012maximum}%
  \BibitemOpen
  \bibfield  {author} {\bibinfo {author} {\bibfnamefont {U.}~\bibnamefont
  {Lucia}},\ }\bibfield  {title} {\bibinfo {title} {Maximum or minimum entropy
  generation for open systems?},\ }\href@noop {} {\bibfield  {journal}
  {\bibinfo  {journal} {Physica A: Statistical Mechanics and its Applications}\
  }\textbf {\bibinfo {volume} {391}},\ \bibinfo {pages} {3392} (\bibinfo {year}
  {2012})}\BibitemShut {NoStop}%
\bibitem [{\citenamefont {Chatterjee}(2016)}]{chatterjee2016thermodynamics}%
  \BibitemOpen
  \bibfield  {author} {\bibinfo {author} {\bibfnamefont {A.}~\bibnamefont
  {Chatterjee}},\ }\bibfield  {title} {\bibinfo {title} {Thermodynamics of
  action and organization in a system},\ }\href@noop {} {\bibfield  {journal}
  {\bibinfo  {journal} {Complexity}\ }\textbf {\bibinfo {volume} {21}},\
  \bibinfo {pages} {307} (\bibinfo {year} {2016})}\BibitemShut {NoStop}%
\bibitem [{\citenamefont {Georgiev}\ and\ \citenamefont
  {Chatterjee}(2016)}]{georgiev2016road}%
  \BibitemOpen
  \bibfield  {author} {\bibinfo {author} {\bibfnamefont {G.~Y.}\ \bibnamefont
  {Georgiev}}\ and\ \bibinfo {author} {\bibfnamefont {A.}~\bibnamefont
  {Chatterjee}},\ }\bibfield  {title} {\bibinfo {title} {The road to a
  measurable quantitative understanding of self-organization and evolution},\
  }in\ \href@noop {} {\emph {\bibinfo {booktitle} {Evolution and Transitions in
  Complexity}}}\ (\bibinfo  {publisher} {Springer},\ \bibinfo {year} {2016})\
  pp.\ \bibinfo {pages} {223--230}\BibitemShut {NoStop}%
\bibitem [{\citenamefont {Landau}\ and\ \citenamefont
  {Lifshitz}(2013)}]{landau2013course}%
  \BibitemOpen
  \bibfield  {author} {\bibinfo {author} {\bibfnamefont {L.~D.}\ \bibnamefont
  {Landau}}\ and\ \bibinfo {author} {\bibfnamefont {E.~M.}\ \bibnamefont
  {Lifshitz}},\ }\href@noop {} {\emph {\bibinfo {title} {Course of theoretical
  physics}}}\ (\bibinfo  {publisher} {Elsevier},\ \bibinfo {year}
  {2013})\BibitemShut {NoStop}%
\bibitem [{\citenamefont {Garc{\'\i}a-Morales}\ \emph
  {et~al.}(2008)\citenamefont {Garc{\'\i}a-Morales}, \citenamefont {Pellicer},\
  and\ \citenamefont {Manzanares}}]{garcia2008thermodynamics}%
  \BibitemOpen
  \bibfield  {author} {\bibinfo {author} {\bibfnamefont {V.}~\bibnamefont
  {Garc{\'\i}a-Morales}}, \bibinfo {author} {\bibfnamefont {J.}~\bibnamefont
  {Pellicer}},\ and\ \bibinfo {author} {\bibfnamefont {J.~A.}\ \bibnamefont
  {Manzanares}},\ }\bibfield  {title} {\bibinfo {title} {Thermodynamics based
  on the principle of least abbreviated action: Entropy production in a network
  of coupled oscillators},\ }\href@noop {} {\bibfield  {journal} {\bibinfo
  {journal} {Annals of Physics}\ }\textbf {\bibinfo {volume} {323}},\ \bibinfo
  {pages} {1844} (\bibinfo {year} {2008})}\BibitemShut {NoStop}%
\bibitem [{\citenamefont {Pauli}\ and\ \citenamefont
  {Enz}(2000)}]{pauli2000thermodynamics}%
  \BibitemOpen
  \bibfield  {author} {\bibinfo {author} {\bibfnamefont {W.}~\bibnamefont
  {Pauli}}\ and\ \bibinfo {author} {\bibfnamefont {C.~P.}\ \bibnamefont
  {Enz}},\ }\href@noop {} {\emph {\bibinfo {title} {Thermodynamics and the
  kinetic theory of gases}}},\ Vol.~\bibinfo {volume} {3}\ (\bibinfo
  {publisher} {Courier Corporation},\ \bibinfo {year} {2000})\BibitemShut
  {NoStop}%
\bibitem [{\citenamefont {Taniguchi}\ and\ \citenamefont
  {Cohen}(2007)}]{taniguchi2007onsager}%
  \BibitemOpen
  \bibfield  {author} {\bibinfo {author} {\bibfnamefont {T.}~\bibnamefont
  {Taniguchi}}\ and\ \bibinfo {author} {\bibfnamefont {E.}~\bibnamefont
  {Cohen}},\ }\bibfield  {title} {\bibinfo {title} {Onsager-machlup theory for
  nonequilibrium steady states and fluctuation theorems},\ }\href@noop {}
  {\bibfield  {journal} {\bibinfo  {journal} {Journal of Statistical Physics}\
  }\textbf {\bibinfo {volume} {126}},\ \bibinfo {pages} {1} (\bibinfo {year}
  {2007})}\BibitemShut {NoStop}%
\bibitem [{\citenamefont {Chatterjee}\ \emph {et~al.}(2019)\citenamefont
  {Chatterjee}, \citenamefont {Yadati}, \citenamefont {Mears},\ and\
  \citenamefont {Iannacchione}}]{chatterjee2019coexisting}%
  \BibitemOpen
  \bibfield  {author} {\bibinfo {author} {\bibfnamefont {A.}~\bibnamefont
  {Chatterjee}}, \bibinfo {author} {\bibfnamefont {Y.}~\bibnamefont {Yadati}},
  \bibinfo {author} {\bibfnamefont {N.}~\bibnamefont {Mears}},\ and\ \bibinfo
  {author} {\bibfnamefont {G.}~\bibnamefont {Iannacchione}},\ }\bibfield
  {title} {\bibinfo {title} {Coexisting ordered states, local equilibrium-like
  domains, and broken ergodicity in a non-turbulent rayleigh-b{\'e}nard
  convection at steady-state},\ }\href@noop {} {\bibfield  {journal} {\bibinfo
  {journal} {Scientific reports}\ }\textbf {\bibinfo {volume} {9}},\ \bibinfo
  {pages} {10615} (\bibinfo {year} {2019})}\BibitemShut {NoStop}%
\bibitem [{\citenamefont {Yadati}\ \emph {et~al.}(2019)\citenamefont {Yadati},
  \citenamefont {Mears},\ and\ \citenamefont {Chatterjee}}]{yadati2019spatio}%
  \BibitemOpen
  \bibfield  {author} {\bibinfo {author} {\bibfnamefont {Y.}~\bibnamefont
  {Yadati}}, \bibinfo {author} {\bibfnamefont {N.}~\bibnamefont {Mears}},\ and\
  \bibinfo {author} {\bibfnamefont {A.}~\bibnamefont {Chatterjee}},\ }\bibfield
   {title} {\bibinfo {title} {Spatio-temporal characterization of thermal
  fluctuations in a non-turbulent rayleigh--b{\'e}nard convection at steady
  state},\ }\href@noop {} {\bibfield  {journal} {\bibinfo  {journal} {Physica
  A: Statistical Mechanics and its Applications}\ ,\ \bibinfo {pages} {123867}}
  (\bibinfo {year} {2019})}\BibitemShut {NoStop}%
\bibitem [{\citenamefont {El~Kaabouchi}\ and\ \citenamefont
  {Wang}(2015)}]{el2015least}%
  \BibitemOpen
  \bibfield  {author} {\bibinfo {author} {\bibfnamefont {A.}~\bibnamefont
  {El~Kaabouchi}}\ and\ \bibinfo {author} {\bibfnamefont {Q.~A.}\ \bibnamefont
  {Wang}},\ }\bibfield  {title} {\bibinfo {title} {Least action principle and
  stochastic motion: a generic derivation of path probability},\ }in\
  \href@noop {} {\emph {\bibinfo {booktitle} {Journal of Physics: Conference
  Series}}},\ Vol.\ \bibinfo {volume} {604}\ (\bibinfo {organization} {IOP
  Publishing},\ \bibinfo {year} {2015})\ p.\ \bibinfo {pages}
  {012011}\BibitemShut {NoStop}%
\bibitem [{\citenamefont {Vilar}\ and\ \citenamefont
  {Rubi}(2001)}]{vilar2001thermodynamics}%
  \BibitemOpen
  \bibfield  {author} {\bibinfo {author} {\bibfnamefont {J.~M.}\ \bibnamefont
  {Vilar}}\ and\ \bibinfo {author} {\bibfnamefont {J.}~\bibnamefont {Rubi}},\
  }\bibfield  {title} {\bibinfo {title} {Thermodynamics “beyond” local
  equilibrium},\ }\href@noop {} {\bibfield  {journal} {\bibinfo  {journal}
  {Proceedings of the National Academy of Sciences}\ }\textbf {\bibinfo
  {volume} {98}},\ \bibinfo {pages} {11081} (\bibinfo {year}
  {2001})}\BibitemShut {NoStop}%
\bibitem [{\citenamefont {Srinivasarao}\ \emph {et~al.}(2019)\citenamefont
  {Srinivasarao}, \citenamefont {Iannacchione},\ and\ \citenamefont
  {Parikh}}]{srinivasarao2019biologically}%
  \BibitemOpen
  \bibfield  {author} {\bibinfo {author} {\bibfnamefont {M.}~\bibnamefont
  {Srinivasarao}}, \bibinfo {author} {\bibfnamefont {G.~S.}\ \bibnamefont
  {Iannacchione}},\ and\ \bibinfo {author} {\bibfnamefont {A.~N.}\ \bibnamefont
  {Parikh}},\ }\bibfield  {title} {\bibinfo {title} {Biologically inspired
  far-from-equilibrium materials},\ }\href@noop {} {\bibfield  {journal}
  {\bibinfo  {journal} {MRS Bulletin}\ }\textbf {\bibinfo {volume} {44}},\
  \bibinfo {pages} {91} (\bibinfo {year} {2019})}\BibitemShut {NoStop}%
\bibitem [{\citenamefont {Chatterjee}\ and\ \citenamefont
  {Iannacchione}(2019)}]{chatterjee2019many}%
  \BibitemOpen
  \bibfield  {author} {\bibinfo {author} {\bibfnamefont {A.}~\bibnamefont
  {Chatterjee}}\ and\ \bibinfo {author} {\bibfnamefont {G.}~\bibnamefont
  {Iannacchione}},\ }\bibfield  {title} {\bibinfo {title} {The many faces of
  far-from-equilibrium thermodynamics: Deterministic chaos, randomness, or
  emergent order?},\ }\href@noop {} {\bibfield  {journal} {\bibinfo  {journal}
  {MRS Bulletin}\ }\textbf {\bibinfo {volume} {44}},\ \bibinfo {pages} {130}
  (\bibinfo {year} {2019})}\BibitemShut {NoStop}%
\bibitem [{\citenamefont {Chatterjee}\ \emph {et~al.}(2020)\citenamefont
  {Chatterjee}, \citenamefont {Mears}, \citenamefont {Yadati},\ and\
  \citenamefont {Iannacchione}}]{chatterjee2019overview}%
  \BibitemOpen
  \bibfield  {author} {\bibinfo {author} {\bibfnamefont {A.}~\bibnamefont
  {Chatterjee}}, \bibinfo {author} {\bibfnamefont {N.}~\bibnamefont {Mears}},
  \bibinfo {author} {\bibfnamefont {Y.}~\bibnamefont {Yadati}},\ and\ \bibinfo
  {author} {\bibfnamefont {G.~S.}\ \bibnamefont {Iannacchione}},\ }\bibfield
  {title} {\bibinfo {title} {An overview of emergent order in
  far-from-equilibrium driven systems: From kuramoto oscillators to
  rayleigh--b{\'e}nard convection},\ }\href@noop {} {\bibfield  {journal}
  {\bibinfo  {journal} {Entropy}\ }\textbf {\bibinfo {volume} {22}},\ \bibinfo
  {pages} {561} (\bibinfo {year} {2020})}\BibitemShut {NoStop}%
\bibitem [{\citenamefont {Far{\'\i}as}\ \emph {et~al.}(2017)\citenamefont
  {Far{\'\i}as}, \citenamefont {Pinto},\ and\ \citenamefont
  {Moya}}]{farias2017temperature}%
  \BibitemOpen
  \bibfield  {author} {\bibinfo {author} {\bibfnamefont {C.}~\bibnamefont
  {Far{\'\i}as}}, \bibinfo {author} {\bibfnamefont {V.~A.}\ \bibnamefont
  {Pinto}},\ and\ \bibinfo {author} {\bibfnamefont {P.~S.}\ \bibnamefont
  {Moya}},\ }\bibfield  {title} {\bibinfo {title} {What is the temperature of a
  moving body?},\ }\href@noop {} {\bibfield  {journal} {\bibinfo  {journal}
  {Scientific reports}\ }\textbf {\bibinfo {volume} {7}},\ \bibinfo {pages} {1}
  (\bibinfo {year} {2017})}\BibitemShut {NoStop}%
\bibitem [{\citenamefont {Mare{\v{s}}}\ \emph {et~al.}(2010)\citenamefont
  {Mare{\v{s}}}, \citenamefont {Hub{\'\i}k}, \citenamefont {{\v{S}}est{\'a}k},
  \citenamefont {{\v{S}}pi{\v{c}}ka}, \citenamefont {Kri{\v{s}}tofik},\ and\
  \citenamefont {St{\'a}vek}}]{marevs2010relativistic}%
  \BibitemOpen
  \bibfield  {author} {\bibinfo {author} {\bibfnamefont {J.}~\bibnamefont
  {Mare{\v{s}}}}, \bibinfo {author} {\bibfnamefont {P.}~\bibnamefont
  {Hub{\'\i}k}}, \bibinfo {author} {\bibfnamefont {J.}~\bibnamefont
  {{\v{S}}est{\'a}k}}, \bibinfo {author} {\bibfnamefont {V.}~\bibnamefont
  {{\v{S}}pi{\v{c}}ka}}, \bibinfo {author} {\bibfnamefont {J.}~\bibnamefont
  {Kri{\v{s}}tofik}},\ and\ \bibinfo {author} {\bibfnamefont {J.}~\bibnamefont
  {St{\'a}vek}},\ }\bibfield  {title} {\bibinfo {title} {Relativistic
  transformation of temperature and mosengeil--ott's antinomy},\ }\href@noop {}
  {\bibfield  {journal} {\bibinfo  {journal} {Physica E: Low-dimensional
  Systems and Nanostructures}\ }\textbf {\bibinfo {volume} {42}},\ \bibinfo
  {pages} {484} (\bibinfo {year} {2010})}\BibitemShut {NoStop}%
\end{thebibliography}
\end{document}